# Privacy Assessment of Software Architectures based on Static Taint Analysis

Marcel von Maltitz, Cornelius Diekmann, and Georg Carle

Technische Universität München
{vonmaltitz,diekmann,carle}@net.in.tum.de

**Abstract.** Privacy analysis is critical but also a time-consuming and tedious task. We present a formalization which eases designing and auditing high-level privacy properties of software architectures. It is incorporated into a larger policy analysis and verification framework and enables the assessment of commonly accepted data protection goals of privacy. The formalization is based on static taint analysis and makes flow and processing of privacy-critical data explicit, globally as well as on the level of individual data subjects. Formally, we show equivalence to traditional label-based information flow security and prove overall soundness of our tool with Isabelle/HOL. We demonstrate applicability in two real-world case studies, thereby uncovering previously unknown violations of privacy constraints in the analyzed software architectures.

## 1 Introduction

Recently, *dynamic taint analysis* [1] has been used successfully in the Android world to enhance user privacy [2, 3]. In this paper, we demonstrate that coarse-grained taint analysis is also applicable to the analysis and auditing of *distributed* architectures, can be done completely *static* (preventing runtime failures), while providing strong *formal guarantees*.

We base our understanding of *privacy* on the concept described by Pfitzmann and Rost [4] and further elaborated on by Bock and Rost [5]. Their proposal has been adapted by the European Union Agency for Network and Information Security (ENISA) [6] and by the German Standardised Data Protection Model [7], showing wide acceptance of their approach.

Analyzing, designing, and auditing distributed software architectures with regard to privacy requirements [4–9] is a complex task: For architecture analysis and design, one requires a specification of privacy goals and a general overview of the data collection, processing, and usage of the software system to be audited. Detailed, often manual, examination is necessary to verify conformity to the specification of an implemented software architecture.

While static label-based information flow is applied in programming languages [10] and a similar *runtime* system has successfully been applied in the Android world [2, 3], for software architectures a clear formalism to express privacy concerns is missing.

As our first contribution, we propose a model based on *static* taint analysis which makes relevant aspects of privacy more tangible. The model generally facilitates the generation of architecture specifications for IT systems and provides guidance for auditing their concrete implementation by directing focus on aspects important for privacy.

We show applicability of our model in two real-world case studies. In the first case study, an energy monitoring system, we could make the informal claims of the system's original architects explicit and verify them. In the second case study, a smartphone measurement framework, we additionally demonstrate the complete audit of the real-world implementation in a fully-automated manner, uncovering previously unknown bugs. To the best of our knowledge, this is the first time that such an audit, which bridges the gap from an abstract taint analysis to complex low-level firewall rules, has been performed completely with the assurance level provided by the theorem prover Isabelle/HOL [11].

Another contribution is integrating a fully formalized privacy model based on taint analysis into a larger, formal policy framework.

While our model is mathematically concise, transforming it into an executable and convenient verification framework is a non-trivial and error-prone task. To provide high confidence in our framework, we have proven all properties of the model and the overall correctness of our framework in Isabelle/HOL (cf. Section Availability).

In detail, our novel contributions to the body of knowledge in this area are:

- A survey of the conceptualization and formal models of privacy (Section 2)
- A formalization of privacy based upon taint analysis, integrated into a fully automated validation and verification framework (Section 4)
- Practical evaluation by two case studies (Section 5)
  - An energy monitoring system
  - A distributed Android measurement framework
- Machine-verifiable proof of correctness in Isabelle/HOL for the presented formalization

This paper is structured as follows: First, we survey related work in Section 2 and compile the requirements for formal assessment of privacy in software architectures in Section 3. We present our formalization in Section 4. Two case studies are presented in Section 5. We critically discuss our results in Section 6 and conclude in Section 7.

## 2 Related Work

### 2.1 Conceptualization of Privacy

The abstract concept of 'security' has been made more tangible and verifiable by deriving protection goals, in particular *confidentiality*, *integrity*, and *availability* (CIA) [12]. Pfitzmann and Rost [4] applied the same methodology to the abstract concept of 'privacy'. They derived the protection goals *unlinkability* (of collected data) and *transparency* (of the data collection and manipulation processes). Unlinkability describes the impossibility to combine present information so that no further information gain can be achieved. Unlinkability itself has already been formalized before by Steinbrecher [13] as a generalization of anonymity. Transparency requires that users can gain insight into processes and software architectures which work with privacy-critical data. It realizes the base upon which surveillance and control of data processing can be carried out. Later, Bock and Rost [5] added *intervenability* (of a data subject in the data processing), effectively creating another triad of protection goals. Intervenability addresses the ability, on the one hand, of data subjects to carry out control over their data and to exercise their user rights and, on the other hand, of process owners to demonstrably be in control of their technical systems

The proposal of Bock, Rost, and Pfitzmann has been adapted in the "Privacy and Data Protection by Design" report [6] and extended by the German Standardized Data Protection Model [7].[1] The German model adds *data minimization* as another protection goal when striving for privacy-preservation. They argue that privacy-threats, being misuse and abuse of data, is most effectively mitigated when the amount of processed data is reduced itself. Hence, data minimization aims for reducing the amount of privacy-critical information from the beginning and in each step where this is possible. They consider this protection goal to be fundamental, as it does not *protect* present vulnerable assets, but *reduces* the actual amount of assets being at risk.

In consequence, when reducing the amount of critical data or the criticality itself, the required strength of data protection measures can also be reduced. This is based on their assumption that the best data protection is not processing privacy-critical data at all.

Another conceptualization of privacy is the *Global Privacy Standard* (GPS) [8]. Comparing "leading privacy practices and codes from around the world", the 27th International Data Protection Commissioners Conference aimed for "develop[ing] a harmonized set of fair information practices". It identified ten principles to be respected when handling personal data of individuals. The principles regard the complete lifecycle of data from collection to processing to final deletion. While being more organizationally than technically oriented, compatibility to the aforementioned approach can be identified. The protection goals of transparency is partially encoded in the principle of *openness*, which allows individuals to have access to "[i]nformation about the policies and practices relating to the management of personal information". *Data minimization* is directly represented but also includes the minimization of data linkability, effectively stating a weaker variant of unlinkability. Lastly, the principle of *compliance* corresponds to intervenability.

In 2010, Cavoukian [9] developed the information management principles, *Privacy by Design* (PbD). PbD identifies principles which help considering privacy-protection from the beginning of the design of a system in order to make it a default property of newly developed systems. For existing systems, the principles give guidance for manual privacy audits. Structurally, PbD is located on a higher level of abstraction than GPS. Some PbD principles subsume multiple GPS statements; further principles are introduced by PbD, which were not yet covered by GPS.

[1] This concept is also designed to be compatible with the data protection laws of Germany.

Hoepman [14] identifies a gap between the high-level principles stated by PbD and their application by incorporating privacy-protection into the design of a concrete technical system. He states that further measures, *privacy design strategies*, must be provided to software engineers and developers in order to actually implement privacy. From a body of current privacy laws and policies, similar to GPS [8], he derived his proposed strategies. Due to the stated motivation, his work hence is more concrete than either GPS or PbD. The first half of the presented strategies is data-centric and describe specific measures in order to reduce the criticality of data present in a system. The second half is process-oriented and focuses the relation of the system to their users. The protection goals initially stated can be rediscovered in Hoepman's work: The strategy *minimize* corresponds to data minimization, *separate* aims for unlinkability, *inform* partially covers transparency, and *control* together with *demonstrate* reflect intervenability.

In line with related work, we also base our understanding of privacy upon the data protection goals of *unlinkability*, *transparency*, *intervenability*, and *data minimization*.

### 2.2 Formal models for privacy

Pfitzmann and Hansen [15], as well as Hoepman [14], contribute groundwork for formalising privacy: They provide definitions of frequent terms in the context of security and privacy. Additionally they present high-level strategies of handling data. Nevertheless, they do not provide a formal approach to privacy themselves.

Steinbrecher and Köpsell [13] developed a valuable generalization of anoymity to unlinkability. They contributed a model of unlinkability based on set partitioning and metrics of relation uncertainty between entities based on information theoretic entropy. While being useful in their own means, their approach addresses only a single aspect of privacy.

Hughes and Shmatikov [16] developed a "modular framework for formalizing properties of computer systems in which an observer has only partial information about system behaviour". Hence, they provide a theoretic tool for creating formalisms which model an adversary's perspective as partial knowledge about a confidential function. In their case study, they regard the security properties anonymity and privacy. Here, they consider privacy to be anonymity of relationships of agents in a system, and explicitly do not cover privacy as protection of personal information. Neither does their approach already allow the assessment of concrete systems.

Fischer-Hübner and Ott [17] define a formal task-based privacy model based on a state mashine model. They aim for deriving formal privacy invariants, expressed as logical predicates, from informal privacy policies. While their work is similar to ours as both approaches consider information flow in systems, their work is orthogonal to ours, as they address the privacy properties of *purpose binding* and *necessity of collection and processing*. They provide an implementation of their model in the Generalized Framework for Access Control Approach in Unix System V.

Most other publications with regard to modeling privacy focus only on a specific subtopic. The proposed models are not generalizable to other areas of application. The most prominent topics are information publishing [18–20], RFID protocols [21–28] and location privacy [29–33].

### 2.3 Evaluation Assurances

Even for clearly specified privacy requirements, the confidence in a software evaluation may vary vastly. For example, the Common Criteria [34] define several Evaluation Assurance Levels (EAL). For the highest assurance level, formal verification is required, e. g. using the theorem prover Isabelle/HOL [11]. One remarkable work in the field of formal verification with Isabelle/HOL is the verification of information flow enforcement for the `C` implementation of the seL4 microkernel [35]. Similarly, to provide high confidence in our results, we have carried out this research completely in Isabelle/HOL. **The proofs are provided in our theory files (cf. Section Availability)**.

### 2.4 Taint Analysis

We introduce the concepts of taint analysis by a simple, fictional example: A house, equipped with a smart meter to measure its energy consumption. The owner also provides location information via her smartphone to allow the system to turn off the lights when she leaves the home. Once every month, the aggregated energy consumption is sent over the internet to the energy provider for billing.

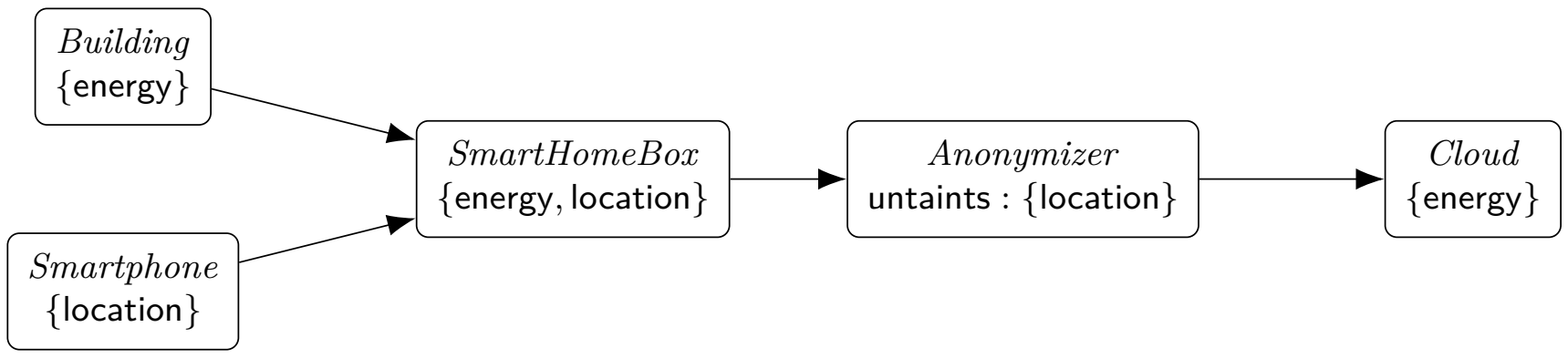

**Fig. 1.** Example: Privacy Concerns and Information Flow in a Smart Home

We are interested in the privacy implications of this setup and perform a taint tracking analysis. The software architecture is visualized in Fig.,1. The *Building* produces information about its energy consumption. Therefore, we label the *Building* as taint source and assign it the energy label. Likewise, the *Smartphone* tracks the location of its owner. Both data is sent to the *SmartHomeBox*. Since the *SmartHomeBox* aggregates all data, it is assigned the set $\{\mathsf{energy}, \mathsf{location}\}$ of taint labels. The user wants to transmit only the energy information, not her location to the energy provider's *Cloud*. Therefore, the *Anonymizer* filters the information and removes all location information. We call this process *untainting*. Let '$\cdot \setminus \cdot$' denote the minus operation on sets. With the *Anonymizer* operating correctly, since $\{\mathsf{energy}, \mathsf{location}\} \setminus \{\mathsf{location}\} = \{\mathsf{energy}\}$, only energy-related information ends up in the energy provider's *Cloud*.

**Taint Analysis in Android** In the Android world, taint analysis has been successfully used recently. TaintDroid [2] uses software instrumentation to track taint information at runtime from pre-defined taint sources, such as the address book or sensors. It then assesses whether tainted data leaves the phone, for example via a network interface. Tracking itself happens on the variable, method, and message level by storing taint labels in memory. Taint labels are also preserved when data is saved to files or databases. TaintDroid's trust model relies on the firmware being the gateway for accessing data considered privacy-critical. Their main contribution is the integration of a multitude of known techniques into a single approach and to demonstrate their applicability to the Android platform.

DroidDisintegrator [3] is built on top of TaintDroid. It first computes a static information flow policy which is presented to the user at the time of app installation. An app is repackaged such that enforcement of this policy at runtime is very lightweight. To compute the policy, DroidDisintegrator relies on dynamic taint analysis and input fuzzing to trigger the valid execution paths of an app. Schuster and Tromer [3] make several important observations: Because of the complexity of information flow security, it is hardly used in mainstream apps. Therefore, Schuster and Tromer simplify the use of information flow security by decreasing the granularity at which DroidDisintegrator operates. They demonstrate that Android *components* provide an appropriate granularity for information flow tracking which "places less burden on the developer than other IFC frameworks" [3, §1.3]. Our model also supports a low-granularity approach by enabling the usage of arbitrary granularity: parts of a software architecture can be modeled with high granularity providing more precision or lower granularity yielding easier applicability. Additionally, we introduce the concept of *system boundaries*, which allows encapsulation of several fine-grained components into one larger coarse-grained component without the loss of information.

Tracking implicit information flow with a practically acceptable false positive/negative rate is still an unsolved problem. DroidDisintegrator avoids this problem in a very domain specific way: Implicit information flow is completely disregarded when the policy is generated. Assuming good programming practice and best-practice use of the Android API, valid and intentional information flow should be explicit. Remaining (probably malicious) implicit information flow, since it is not contained in the policy, is consequently prohibited at runtime. Our generic analysis framework is built on the same principle.

**Further Successful Formal Security Approaches for Android** In a broader context, Bagheri et al. [36] analyzed the Android permission system. They used bounded verification with the Alloy tool to find design flaws. Permissions, which focus on access control, may be considered complementary to taint analysis, which focuses on information flow.

Fragkaki et al. [37] build a model of the Android permission system in order to specify desired security properties and to verify whether these requirements hold for the given system. They show that these requirements are only partly fulfilled by the examined versions of Android. In line with our work, they consider taint analysis for information flow security. They utilize a noninterference model to formalize it in a dynamic context.

### 2.5 Models, Formalization, and *topoS*

"The architecture defines the structure of a software system in terms of components and (allowed) dependencies" [38]. We will stick to this high-level, abstract, implementation-agnostic definition for the formalization. Our case study will evaluate a distributed system as well as an implementation which relies on a component architecture of just one system.

As illustrated in Figure 1, a graph can be conveniently used to describe a system architecture. Since a graph (without taint label annotations) specifies the permitted information flows and all allowed accesses, we will call such a graph a *policy*. We will always assume that a graph $G = (V, E)$ is syntactically well-formed, i. e. all vertices occuring in the edges $E$ are also listed in the nodes $V$.

To analyze, formalize, and verify policies represented as graphs, we utilize the *topoS* [39,40] framework. *topoS* allows specification of predicates over the policy. These predicates are called *security invariants*. The security invariants follow special design criteria to ensure the overall soundness of *topoS*. To define a new security invariant, *topoS* imposes strict proof obligations. In return, *topoS* offers arbitrary composability of all security invariants and generic analysis/verification algorithms.

## 3 Requirement Analysis

Our understanding of privacy focuses on the data protection goals of unlinkability, transparency, intervenability, and data minimization. We require that our approach supports the assessment of a software architecture with respect to privacy by making the handling of privacy-critical data streams explicit and providing insights into the fulfillment of these very protection goals.

### 3.1 Data Privacy Protection Goals

To achieve this overall goal, the findings of the modelling must support the assessment of the following aspects of privacy-preservation in software architectures:

**Unlinkability** When different types of originally unrelated information are processed at the very same realm (e.g. components), then, unlinkability is jeopardized as these components can possibly derive further information by combining the available data (e.g. using operations like set intersection or joining). Hence, we require that our model makes these realms explicit. This is a critical part of the assessment of unlinkability and enables developing remedies for this type of privacy threat.

For improving unlinkability, our model should furthermore give insights how existing data flows can be split in order to avoid realms with the ability to link.

**Transparency and Intervenability** Intervenability is based on knowledge about a system. We require, that the model empowers the stakeholders to improve privacy by making relevant data-flow aspects of privacy-preservation in a given architecture explicit. Data generation, handling, processing and storage must become explicit in the formalism.

Transparency should especially be given for two groups of stakeholders: First, the system providers and administrators must be able to gain an overview over the complete data processing in the given software. This enables them to identify components which unintentionally retrieve and process a certain type of privacy-critical information. Second, our approach should be usable to enable users to gain insights in the processing of their very data. Hence, we require it to be possible to derive "views" on the model which express the specific privacy-implications for a given user.

**Data Minimization** Data minimization targets the reduction of the amount of data flowing through a system (without loss of functionality). Hence, our approach must improve the assessment by providing insights in the amount of relevant data flowing through a software system. This should build an information foundation which makes the constructions of metrics possible. These metrics can enable the estimation of the criticality of the software in general and the criticality of certain components in particular. Similar to requirement unlinkability, the model should support identifying hot spots, where the reduction of the amount of data processing has the biggest positive effect for privacy.

### 3.2 Security and Risk Assessment

Besides support in the assessment of privacy in particular, our approach should support accompanying aspects of security and risk assessment. These are *asset identification*, the creation of a *trust model* and an *adversary model*.

**Asset Identification** We consider two types of assets to the relevant. The data being handled constitutes the primary asset to be protected. The requirement to make this aspect of the system explicit has already been stated in the previous Section 3.1 .

Additionally, components handling the data also represent assets. Hence, we require that our model supports the identification of these assets as well as the assessment of their criticality. A foundation for the latter assessment is the amount of data flows being processed by a given component.

**Trust Model** When handling privacy-critical data, stakeholders have to put confidence in software or some of its components to be trustworthy. The model should make clear, which component is trusted with regard to specific privacy-critical information. Trust here means the confidence that the component works as specified, correctly processes the given type of data and does not leak or retain information for usage of secondary purposes. A relevant special case of trusted behavior is the application of privacy-enhancing technologies. This processing has considerable impact on the trust of each following component, effectively reducing the amount of necessary trust as the data processed became less critical. Hence, we want to have the positive consequences of this type of processing explicitly representable in our formalism.

**Adversary Model** In order to specify adversaries intercepting specific channels between individual systems or compromising a set of systems, it is important not to sacrifice the notion of individual systems in the formalism by only modelling interconnected components without differentiation depending on their locality. Instead, explicitly modelling systems as a group of components allows the assessment of scenarios where single systems are compromised and where eavesdropping on channels between systems is considered. Consequently, the formalism must support combining multiple components into individual systems.

### 3.3 Implementation support

We want, that our formalism not only gives insights into the architecture of a system but also to propose improvements, when privacy constraints are not fulfilled. We will show that our approach enables *topoS* to automatically derive the strictest set of firewall rules, which allows interconnected systems to communicate as specified and suppresses all further communication including all privacy-violating exchanges.

## 4 Formalization & Implementation

We formalize tainting as a security invariant for *topoS*. To foster intuition, we first present a simplified model which does not support trust or untainting. However, we have aligned the paper constructively such that all the results obtained for simple model follow analogously for the full model. All details can be found in our accompanying Isabelle/HOL formalization (cf. Section Availability).

We assume we have a total function $t$ which returns the taint labels for an entity. For example, $t\ SmartHomeBox = \{\mathsf{energy}, \mathsf{location}\}$.

Intuitively, information flow security according to the taint model can be understood as follows. Information leaving a node $v$ is tainted with $v$'s taint labels, hence every receiver $r$ must have the respective taint labels to receive the information. In other words, for every node $v$ in the policy, all nodes $r$ which are reachable from $v$ must have at least $v$'s taint labels. Representing reachability by the transitive closure (i.e. $E^+$), the invariant can be formalized as follows:

$$\textsf{tainting}\ (V, E)\ t \equiv \forall v \in V.\ \forall r \in \{r.\ (v, r) \in E^+\}.\ \ t\ v \subseteq t\ r$$

For this formalization, we discharged the proof obligations imposed by *topoS*. First, the security invariant is monotonic, which means that prohibiting more flows will never make the policy less secure. Second, in case of a violation, there is always a set of flows which are responsible for the violation and the violation can be repaired by prohibiting said flows. We consider tainting as an information flow security (IFS) invariant. For *topoS*, this means a violation occurs as soon as a labeled information reaches an unintended receiver. Our formalization also discharges the additional proof obligations imposed by *topoS* for IFS invariants: A user will unlikely provide a *total* assignment $t$ of taint labels. *topoS* can take a partial assignment and, with the help of a secure default parameter, auto-complete it to a total function. For this, *topoS* imposes the proof obligation for IFS invariants that the default parameter can never hide a violating information flow. In addition, it requires that the default parameter is uniquely defined, i.e. it is the only value which can always uncover violations. Intuitively, if we assume that a user has given labels to the important taint sources, the default parameter needs to be the empty set of taint labels since this will uncover all undesirable flows from labeled to unlabeled sources. Therefore, our invariant fulfills all proof obligations of *topoS*.[2]

## 4.1 Analysis: Tainting vs. Bell-LaPadula Model

The Bell-LaPadula model (BLP) [41, 42] is the traditional, de-facto standard model for label-based information flow security. The question arises whether we can justify our taint model using BLP.

*topoS* comes with a pre-defined formalization of the BLP model [40]. The labels in BLP, often called security clearances, are defined as a total order: unclassified $\leq$ confidential $\leq$ secret $\leq$ topsecret $\leq \ldots$ Let $sc$ be a total function which assigns a security clearance to each node. Since our policy model does not distinguish read from write actions, the BLP invariant simply states that receivers must have the necessary security clearance for the information they receive:

$$\textsf{blp}\ (V, E)\ sc \equiv \forall (v_1, v_2) \in E.\ sc\ v_1 \leq sc\ v_2$$

Inspired by BLP, we show an alternative definition for our tainting invariant:

[2] We only have sketched the rough idea here, the full proofs can be found in our formalization: `interpretation Taints: SecurityInvariant-IFS`

**Lemma 1 (Localized Definition of Tainting).**

$$\mathsf{tainting}\ (V, E)\ t = \forall (v_1, v_2) \in E.\ \ t\ v_1 \subseteq t\ v_2$$

*Proof.* We assume a syntactically well-formed graph. First, we note that the tainting invariant can be rewritten such that instead of quantifying over all vertices, it quantifies over the first node of all edges. Subsequently, by induction over the transitive closure, the invariant can be rewritten to the desired form.

Lemma 1 also provides a computational efficient formula, which only iterates over all edges and never needs to compute a transitive closure.

We will now show that one tainting invariant is equal to BLP invariants for every taint label. We define a function project $a$ $Ts$, which translates a set of taint labels $Ts$ to a security clearance depending on whether $a$ is in the set of taint labels. Formally, project $a\ Ts \equiv$ **if** $a \in Ts$ **then** confidential **else** unclassified. Using function composition, the term *project* $a \circ t$ is a function which first looks up the taint labels of a node and projects them afterwards.

**Theorem 1 (Tainting and Bell-LaPadula Equivalence).**

$$\mathsf{tainting}\ G\ t \longleftrightarrow \forall a.\ \mathsf{blp}\ G\ (\mathsf{project}\ a\ \circ\ t)$$

*Proof.* See `Analysis_Tainting.thy`

The '→'-direction of our theorem shows that one tainting invariant guarantees individual privacy according to BLP for each taint label. This implies that every user of a software can obtain her personal privacy guarantees. This fulfills the *transparency* requirement for individual users.

The '←'-direction shows that tainting is as expressive as BLP. This justifies the theoretic foundations w.r.t. the well-studied BLP model. These findings are in line with Denning's lattice interpretation [43]; however, to the best of our knowledge, we are the first to discover and formally prove this connection in the presented context.

The theorem can be generalized for arbitrary (but finite) sets of taint labels $A$. The project function then maps to a numeric value of a security clearance by taking the cardinality of the intersection of $A$ with $Ts$. For example, if we want to project {location, temp}, then {name} is unclassified, {name, location, zodiac} is confidential, and {name, location, zodiac, temp} is secret.

### 4.2 Untainting and Adding Trust

Real-world application requires the need to untaint information, for example, when data is encrypted or properly anonymized. The taint labels now consist of two components: the labels a node taints and the labels it untaints. Let $t$ be a total function $t$ which returns the taints and untaints for an entity. We extend the simple tainting invariant to support untainting:

$$\mathsf{tainting}'\ (V, E)\ t \equiv \forall (v_1, v_2) \in E.\ \ \mathsf{taints}\ (t\ v_1) \setminus \mathsf{untaints}\ (t\ v_1) \subseteq \mathsf{taints}\ (t\ v_2)$$

To abbreviate a node's labels, we will write $X$—$Y$, where $X$ corresponds to the $\mathsf{taints}$ and $Y$ corresponds to the $\mathsf{untaints}$. For example, in Fig. 1 we have $t\ \mathit{Anonymizer} = \{\mathsf{energy}\}$—$\{\mathsf{location}\}$.

We impose the type constraint that $Y \subseteq X$, i.e. $\mathsf{untaints} \subseteq \mathsf{taints}$. We implemented the datatype such that $X$—$Y$ is extended to $X \cup Y$—$Y$. Regarding Fig. 1, this merely appears to be a very convenient abbreviation. In particular, $t\ \mathit{Anonymizer}$ now corresponds to $\{\mathsf{energy}, \mathsf{location}\}$—$\{\mathsf{location}\}$, for which the $\mathsf{tainting}'$ invariant holds and which also corresponds to our intuitive understanding of untainting. However, this is a fundamental requirement for the overall soundness of the invariant. Without the type constraint, there can be dead $\mathsf{untaints}$, i.e. $\mathsf{untaints}$ which can never have any effect, which would violate the uniqueness property required by *topoS* for default parameters and cause further problems in pathological corner cases. Yet, with this type constraint, as indicated earlier, all insights obtained for the simple model now follow analogously for this model.

*Analysis: Tainting' vs. Bell-LaPadula Model'* In *topoS*' library, there is a predefined formalization of the BLP model with trusted entities [40]. In the context of BLP, a trusted entity is allowed to declassify information, i.e. receive information of any security clearance and redistribute with its own clearance (which may be lower than the clearance of the received information). This concept is comparable to untainting. Let $\mathsf{trusted}$ extract the trusted flag from an entity's attributes and $\mathsf{c}$ extract the security clearance.

Our insights about the equality follow analogously. Let $\mathsf{project}\ a\ (X$—$Y)$ be a function which translates $\mathsf{taints}$ $(X)$ and $\mathsf{untaints}$ $(Y)$ labels to security clearances $\mathsf{c}$ and the $\mathsf{trusted}$ flag. Let $\mathsf{c} = \mathbf{if}\ a \in (X \setminus Y)\ \mathbf{then}\ \mathsf{confidential}\ \mathbf{else}\ \mathsf{unclassified}$ and $\mathsf{trusted} = a \in Y$.

**Theorem 2 (Tainting and Bell-LaPadula Equivalence – full).**

$$\mathsf{tainting}'\ G\ t \longleftrightarrow \forall a.\ \mathsf{blp}'\ G\ (\mathsf{project}\ a \circ t)$$

Similarly to the version without trust, the theorem can be generalized for arbitrary (but finite) sets of taint labels.

### 4.3 System Boundaries

*topoS* provides a number of useful analyses. For example, given a taint label specification, *topoS* can compute all permitted flows which is invaluable for validating a given specification. However, *topoS* might lack knowledge about architectural constraints which leads to the computation of an unrealistic amount of flows. For example, in Fig. 3, *CollectDroid* is one physical machine and *Dec-A* is one isolated software component running on it. *Dec-A* is neither physically reachable from outside of the machine nor has direct network connectivity itself. We want to provide this knowledge to *topoS*. Therefore, we model systems with clearly defined boundaries. This is visualized in Fig. 3 by the dotted rectangles; entities which are partially outside the rectangle represent system boundaries. We define *internal* components (such as *Dec-A*) as nodes which are only accessible from inside the system. We define *passive system boundaries* to be boundaries which only accept incoming connections. Analogously, *active system boundaries* are boundaries which only establish outgoing connections. A *boundary* may be both.

A *topoS* invariant must either be an access control or information flow invariant. An access control invariant restricts accesses *from* the outside, an information flow invariant restricts leakage *to* the outside. However, internal components of a system, e. g. *Dec-A*, require both: they should neither be accessible from components outside of the system nor leak data to outside components. We overcame this limitation of *topoS* by constructing a model for system boundaries which is internally translated to two invariants: an access control invariant and an information flow invariant. We have integrated the concept of system boundaries into *topoS* and proven the desired behavior of our implementation.

## 5 Evaluation

For the purpose of a case study, we use two distributed systems for data collection which are deployed at the Technical University of Munich. First, we describe their architectures, their purpose, and their handling of privacy-critical data as well as their architecture. Then we present the modeling of the architectures in *topoS*. In the second case study, the steps of architecture and taint label specification are performed analogous to the first study and hence are not further elaborated on. In a last step, we present the results of the insights gained by the application of our formalism.

### 5.1 IDEM

The project IDEM [44] focuses on energy monitoring systems (EMS). The purpose of an EMS is to support the reduction of energy consumption in the monitored area by providing detailed and fine-grained insights in the actual use of energy per room. These information help to identify the most greedy, faulty, or incorrectly configured devices in the area and enable responsible staff to take appropriate actions like repairing or replacing identified devices. Furthermore, an EMS can also be used to carry out consumption billing with higher granularity.

The hardware of an off-the-shelf EMS consists of two components. A logging unit which is attached to the fuse box of the area to monitor and transforms the analogue signal into digital data points. A directly connected controller obtains these data points and POSTs them via HTTPS to a cloud service. Data is stored there and made available via a web-based GUI providing measures for statistical analysis and visualization.

**Privacy Criticality** We consider the measured information to be privacy-critical [44], as the energy consumption of devices is highly correlated with the presence and behaviour of building inhabitants using these devices. We exemplify this point with the following scenario: Given, the measured rooms are office rooms for at maximum two person each. Then, each room contains approximately the same equipment like computer, display, telephone, etc. The typical use of these devices results in a certain consumption pattern on every working day. E.g. due to automatic standby functions of displays as well as computers and similar effects, power consumption lowers when people leave their working place. This allows to carry out surveillance of user behaviour by misusing the EMS. Especially tracking of work begin and end times as well as the number and duration of breaks becomes easily possible. We modified the system in order to achieve the following improvements: Collected information does not have to leave the building but is stored on a local server. Access control is not carried out by user authentication against a trusted system which has full access but by cryptographic means instead. Data is preprocessed for different audiences, so that each audience only gets the minimum of data which is needed to carry out their task. From the provider's point of view, these measures are motivated by data protection laws and policies as well as the mitigation of data breach consequences.

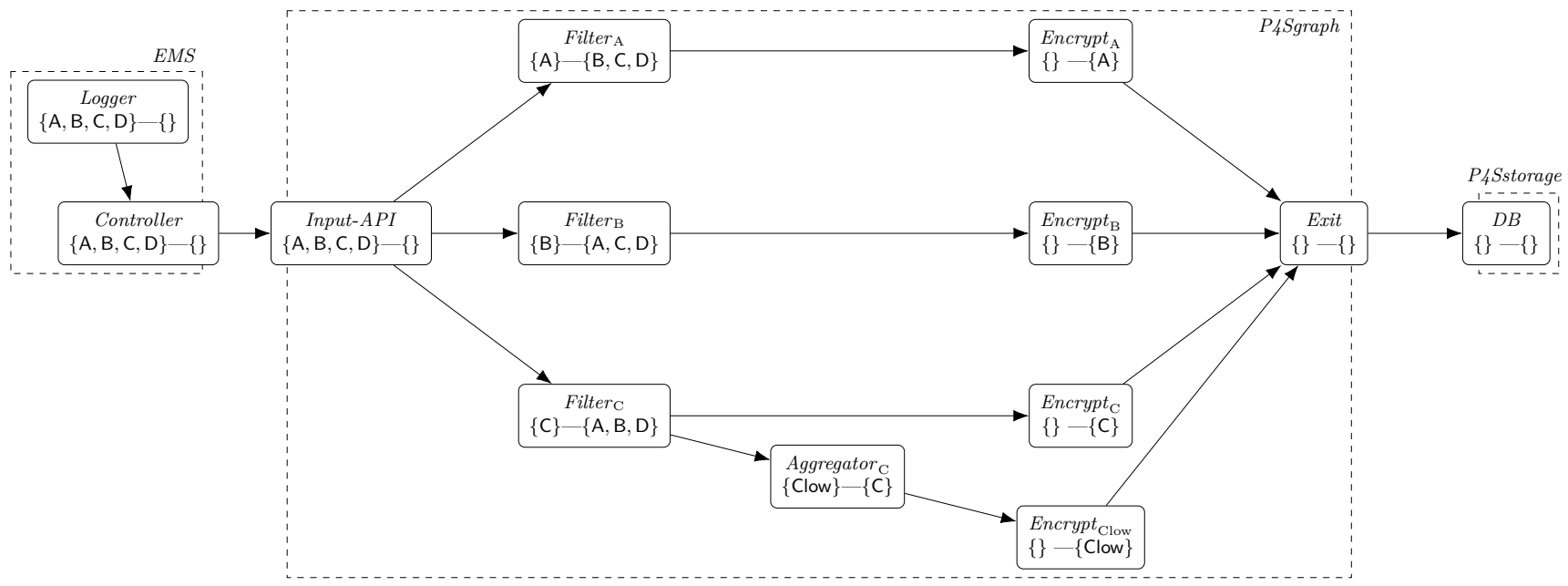

**Fig. 2.** IDEM Architecture

**Architecture Specification** The EMS can be modelled by two components: The *logger*, being the data source, creates digital data points and the *controller* pushes those data points to a given target. The controller is reconfigured not to send data to the cloud but to the local privacy-preserving preprocessing and storage system (P4S) via a secured channel. Hence, the logger is an internal component while the controller is an active system boundary. The analogue/digital transformation is not considered a relevant border in our context.

The P4S consists of the Input-API, a directed acyclic graph of preprocessing modules, and a storage system. The Input-API obtains the data, when POSTed to the P4S, and hence is a passive system boundary. The Input-API acts as the entry node of the graph structure. The exit node forwards the preprocessed data to the storage component, consequently being an active system boundary. All other nodes are regarded to be internal, but are also considered in the modelling. The actual shape of the graph and the selection of nodes depend on the concrete system goals and the user roles which shall later obtain the collected information. The purpose of the other nodes is as follows: Given a predicate, *filtering nodes* only pass through data points which match the predicate, all other data is dropped. The aim is to split a single data stream containing data points from all rooms and users into multiple streams which only contain data about a single entity. It is obvious that the former combined stream is more privacy-critical as correlations between different entities can be analysed. Separation itself is already a measure of privacy improvement [14].

*Aggregation nodes* transform multiple data points into a single data point using aggregate functions like the sum, average, or median. Aggregation is also a measure of privacy improvement as it removes information details and prevents propagation of data which has higher information value and accuracy than needed for a given purpose [14]. An example where this measure can be taken, is when billing information is derived from the consumption data which initially had a precision of seconds to minutes. This data stream can be reduced to data points which only represent the overall amount consumed per month.

*Encryption nodes* implement access control on the data level. Each of these nodes encrypt the incoming data for a set of preconfigured recipients. The privacy-critical data stream is then protected until its decryption by one of the authorized recipients. Besides privacy-uncritical publicly published information, all information flowing on a path from the source of the graph to the data storage component has to pass an encryption node before being forwarded to the latter. Each path is effectively the preprocessing steps for a specific group of recipients.

After flowing through the exit node, the data is transmitted to and saved in the storage component of the P4S. As data is encrypted, access is enabled by indexes build upon stream meta data. This includes e.g. the data type and timestamps or time intervals. Decryption happens on the data consumers' local system after querying and retrieval.

**Taint Label Specification** We specify the taint labels as described in Fig. 2. For the current scenario, we assume four users, each possessing their own taint label $\mathsf{A}$, $\mathsf{B}$, $\mathsf{C}$, and $\mathsf{D}$.

The logger is a central unit, hence it taints all labels $\{\mathsf{A}, \mathsf{B}, \mathsf{C}, \mathsf{D}\}$. The data points are internally forwarded to the controller, requiring the same taint labels for it. The same applies for forwarding them to the Input-API, although traversing system boundaries.

Filtering nodes remove all but one specified taint label. This is expressed in the combination of tainting and untainting labels. Each filtering node $F_K$ with $K \in M$ obtains data of all taints and then removes the data of the taints $M \setminus K$, effectively untainting the labels $M \setminus K$. Optionally, as shown for user $\mathsf{C}$, an ag-gregation node may untaint given data and derive a new *low* taint label from it. This reflects two aspects of its behaviour: First, it receives privacy-critical data of only a single taint. Second, the output data of taint $K_{\text{low}}$ is considered to be semantically less privacy-critical than the input data with taint $K$. Encryption nodes untaint a given label by encrypting the information. While the input data can possess arbitrary privacy-criticality, the output data is always considered completely uncritical[3]. The encryption nodes allow that each following component does not need any taints in order to receive, store, and forward the data. This especially includes in this scenario the exit node, the P4S storage component, and the query API. On decryption on the local system of the authorized recipients the data restores its previous privacy-criticality.

*Remark:* Often, it is more sensible to model aggregation as *reduction* of privacy-criticality instead of *complete removal*. Formally, this can be represented by replacing an initial taint label with a new one, while implying that the latter is less critical than the former. Similarly, it is imaginable that merge nodes exist which combine data of several different taint labels to a combined stream of greater privacy-criticality, analogously indicated by a more critical taint label. Currently, the formalism does not support encoding the rating of criticality explicitly, as taint labels are only categorical values without any means of comparison. In future work, we will add the ability to define a partial order of taint labels to our model to allow better assessment of criticality.

**Results and Insights** In the following we describe the results of the formalization of the architecture at hand.

*Data Privacy Protection Goals* *Data minimization* is performed relatively late. Data is not collected locally with regard to individual users but rather in a centralized manner. As a consequence, central components have full access to the privacy-critical data of all users. More importantly, critical data crosses system boundaries, which makes them susceptible to network attacks. However, an advantage of the design is that data is untainted by encryption before storage. This prevents attacks of the database which would jeopardize the whole history of collected data.

By using the model, it becomes *transparent* that the EMS is critical for all users. Furthermore, exactly one path in the P4S graph is dedicated to each user; the untainting is completely identical for all users. Protection of stored data also becomes visible. The data from user C is further processed and derived values are also stored in the database. If user C is not aware of that process, she can *intervene* and post an application of information to gain further insights.

Due to the aforementioned violations of locality and separation principles, data is *linkable* in the EMS and in the P4S graph before separation by filters.[4]

[3] Under the common assumption that strong state of the art encryption is utilized.

[4] The actual instantiation of the model influences the linkability of data. When multiple aggregation nodes for *different* taints are actually implemented as the *very same* component, this component possesses the ability to link while this does not become

While not included in the current case study, if combined flows e.g. of several users existed, exactly these links would also become visible in the model.

*Security and Risk Assessment* The first *asset* with regard to privacy-protection is the EMS. The EMS is fully privacy-critical and does not employ any protection and untainting measures. Given that it is not build but bought, protection might only be possible by wrapping it. The next critical part of the architecture is the connection between the EMS controller and the API of the P4S. If applicable, TLS protection can be employed, otherwise measures which do no need support on the side of the EMS are necessary. Furthermore, the P4S is considered to be privacy-critical as it obtains the full stream of data.

On the level of components, from source to sink, the P4S constitutes a gradient with respect to privacy-criticality. With each step, privacy-critical data is more and more untainted. At last, data in the database is completely untainted; hence, further measures of protection are not strictly necessary here.

We *trust* in the integrity of the EMS, that is, that it will only send its data to the P4S. Accordingly, we trust the connection not to be susceptible to eavesdropping or man-in-the-middle attacks. Moreover, we expect that there are no data leaks in the P4S and data is only transmitted to the database. There is, however, no need to trust the database itself.

With regard to untainting, we expect the correctness of privacy-protection measures implemented in the filtering, aggregation, and encryption nodes. Without them, components which are now deemed not to process any privacy-relevant data also become critical. Especially $Aggregator_{\mathrm{C}}$ is a valuable target for further investigation. Here, actual untainting depends more on the very semantics of the applied function than when using, for example, encryption which works without regard to the meaning of the encrypted data.

Based on the trust assumptions and the assessment of criticality, valuable targets for *adversaries* are the EMS and the P4S graph. The EMS has an active boundary, hence, intrusion must be considered, being an attack on the integrity of the component, There are, however, no query APIs which have to be protected against unauthorized data access. The Input-API of the P4S graph is a passive boundary, accepting but not providing data. This interface is not to be considered privacy-critical as it only receives information. Nevertheless, possible attacks include POSTing faulty data and provoking denial of service by flooding the system with data. Also having an active boundary, the exit node, intrusion is a concern like for the EMS.

The exit node is not to be considered critical, due to two reasons: It is an active boundary, not allowing data querying and the data is already encrypted and therefore protected.

The database would be critical as it holds all data yet collected. However, this is also not the case due to the previously employed encryption.

---

visible in the model. An improvement on the implementation side is to enforce complete statelessness of these components.

### 5.2 MeasrDroid

MeasrDroid [45] is a system for collecting smartphone sensor data. The goal is to utilize the gathered information for research purposes. It may collect information about the current battery power, smartphone orientation, properties of surrounding wifi and cellphone networks such as signal strength, latency, and reliability. All information is collected in combination with the current location of the smartphone. This allows to generate maps with respect to the measured properties.

For this purpose, an app is installed on each participating smartphone. The app regularly reads all available (hardware and software) sensor data and encrypts them with a predefined public key. The data is then send to *UploadDroid*, an upload gateway which temporarily stores the collected information. With lower frequency, *CollectDroid*, a trusted database server, pulls the information from *UploadDroid* and decrypts them locally. The data is then stored in a local database on *CollectDroid*. From these information, statistics and graphs are generated and further analyzed.

**Privacy Criticality** Privacy criticality comes from the fact that all information is not only mappable to an individual identifier for each smartphone, but it is also collected including a precise GPS location of it. This enables tracking of users over the time.

**Architecture Specification** Data is repeatedly collected on the smartphone using the dedicated MeasrDroid app and serialized to a JSON string. The sensor component of the app is the data source. As in IDEM, it is not considered to be a boundary in the formal model. The data is directly and locally encrypted by an internal component using the public key of the trusted *CollectDroid* server and not saved on the phone in plaintext.[5] From there it is internally forwarded to the upload component which is an active system boundary and POSTs the data to the *UploadDroid* server.

The following components constitute the MeasrDroid backend infrastructure.

The *CollectDroid* server shall not be accessible from the outside in order to minimize its attack surface. Due to this reason, is it not desired that information is directly pushed to it. Alternatively, the gateway server *UploadDroid* is provided, to which the information can be pushed. Encrypted, and hence privacy-protected information is only temporarily deposited on this server. This server is simultaneously a passive system boundary for the smartphones pushing their data as well as for the *CollectDroid* server pulling the data. Complementary, *CollectDroid* has an active boundary for querying for new data.

The *CollectDroid* server consists of a data collection component, a data decryption component, and an actual storage component. Regularly, the trusted and isolated *CollectDroid* connects to the *UploadDroid* via SSH and pulls the newly received measurement data. It is the vital requirement that it may not be contacted from the outside. This can be modeled by not providing any passive system boundaries. After the transfer, data is decrypted and stored there.

[5] In consequence, even when the phone is stolen, an advisory cannot reconstruct the whole measurement and location history of this phone.

In other words, the competing properties of *outside access* for data pushing on the one hand and *trustworthiness* due to the handling of unencrypted data on the other hand have been separated into two systems.

This derived model is shown in Figure 3.

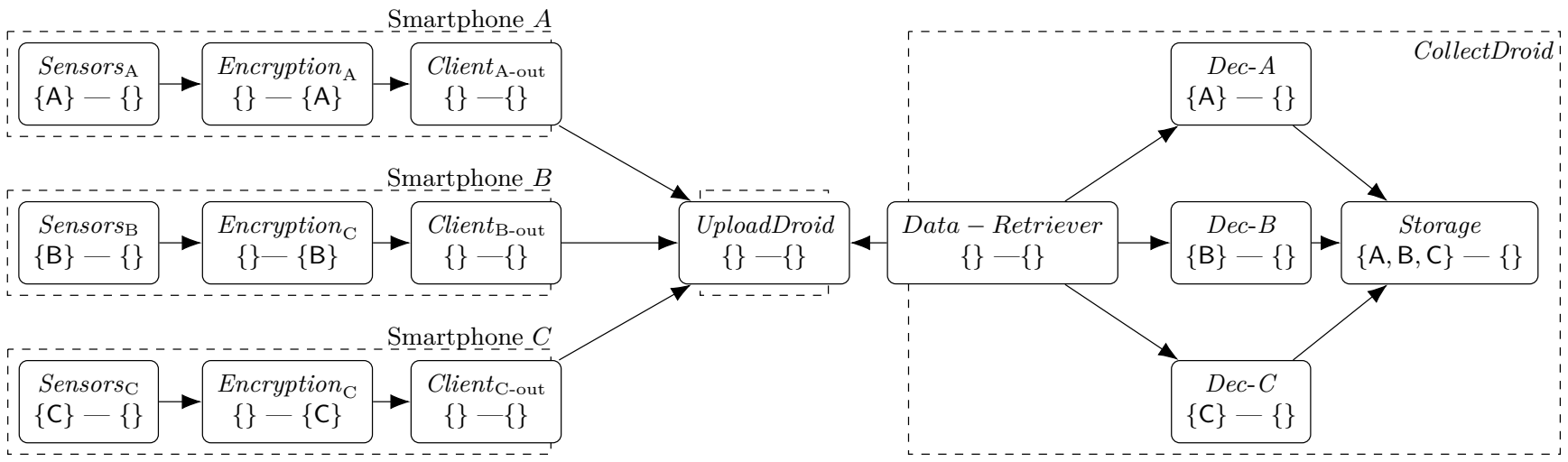

**Fig. 3.** MeasrDroid Architecture

The main purpose of the collected data is to support scientific work. The process of data analysis is carried out manually, hence there exists no further infrastructure from this point on which is relevant for the current elucidations.

**Taint Label Specification** For the current scenario, we assume several users, each possessing their own tainting label. For three users, the tainting labels $\mathsf{A}$, $\mathsf{B}$, and $\mathsf{C}$ exist. Their smartphones collect individual data about them. Hence, the collection component of every smartphone taints the gathered information with the label of the corresponding user. Immediately, the following encryption component again untaints the owner's label. In consequence, the upload component does not have taint labels.

On the side of the MeasrDroid system, the *UploadDroid* only handles encrypted data which it cannot decrypt. Hence, it does not have any taint labels. This also applies for the *CollectDroid*'s data collection component. The *CollectDroid* decrypts the data and has access to all information of every user. Hence, it is necessary to allow the full taint label set, here being $\{\mathsf{A}, \mathsf{B}, \mathsf{C}\}$ for its decryption component (necessarily modelled as three individual nodes to correctly reapply the decrypted data's taints), as well as the storage component.

*Remark:* The current case showed that encryption can completely untaint given information, regardless of their previous taints. For every encryption function, a related decryption function must exist, which restores the previous information. Formally, also all previously applied taint labels are restored. Besides explicit specification of taint labels, it should be possible to relate a pair of encryption and decryption components to enable automatic taint label derivation for the decryption component. The main problem to be avoided is the manual introduction of inconsistencies by updating the taint specificiations of only one entity in such a pair. We created a wrapper function, which creates an taint-untaint-tuple for the encryption as well as the decryption component when providing the corresponding taint and untaint labels as parameters. This solved the aforementioned problem by the mere introduction of syntactic sugar, hence, not requiring changes of the previously specified formalism nor requiring further proofs.

**Results and Insights** Our evaluation of the architecture specification is analogue to the previous case study. Remarkable differences are as follows.

*Data Privacy Protection Goals* Data collection is distributed and separated from the beginning as each smartphone only collects data about its own user.

This has several benefits: *Data minimization* can already be performed by the client application on the smartphone before it crosses any system boundary. If the client-side application is open source, full *transparency* of the running code is given. Also, the foundation for more *intervenability* is laid, as the app can provide measures to the user, to exactly specifiy which information may be collected. Based on separation, *linkability* of different users data is also prevented. Only after storage, data can potentially be linked.

*Security and Risk Assessment* For each user, two assets exist: The user's individual smartphone must be considered critical as well as *CollectDroid*, which is critical for all users. *UploadDroid* is considered completely uncritical.

The smartphone app is trusted that its sensor component only collects legitimate data and that the encryption component performs untainting correctly. Furthermore it is expected that its upload component only pushes this data to *UploadDroid* (if the data would be sent to a different entity, the data cannot be decrypted). Lastly, *CollectDroid* is trusted with respect to data handling.

Depending on the motivation, two types of attacks can be derived: If surveillance of a single individual is desired, the corresponding smartphone as well as the *CollectDroid* are potential attack targets. Hence, the integrity of the smartphone, the app, and the trusted backend components has to be assured. When access to all critical information, e. g. for untargeted correlation purposes is desired, then the trusted backend systems (decryption component and the storage component) become main targets. *CollectDroid* only has an active boundary; intrusion has to be considered as an important attack vector. Consequently a vital protection is a firewall restricting incoming connections.

While it is obvious that the storage component is highly critical and therefore has to be protected, we consider it a valuable hint by *topoS* that the decryption component itself have the same level of criticality and need to be protected with the same amount of diligence.

*Implementation Support* The *CollectDroid* server has no passive boundaries but only active boundaries to obtain data from the *UploadDroid*. It can be completely isolated as mentioned before. This could be reflected in the following firewall configuration:

- Drop all incoming connection requests[6]
- Only allow outgoing connections to *UploadDroid*

*topoS* shows that the *UploadDroid* is a passive system which has to be protected by other means. For the Internet-facing side, each individual user can be given a cryptographic identifier in order to authenticate against the APIs of the mentioned systems. Internally, certificates can be utilized in order to perform mutual authentication of the servers.

### 5.3 Auditing the Real MeasrDroid

The previous sections presented a theoretical evaluation of the architecture of MeasrDroid and consequently provide a recipe for evaluating and auditing the real system. MeasrDroid is deployed and in productive use since 2013. The theoretic evaluation of its architecture, presented in this paper, was not available at the time MeasrDroid was developed. In this section, together with the authors of MeasrDroid we will evaluate the real MeasrDroid implementation with regard to our findings of the previous sections.

First, we collected all physical and virtual machines which are associated with MeasrDroid. We found the following machines:

**droid0** Virtual machine
- IPv4: 131.159.15.16
- IPv6: 2001:4ca0:2001:13:216:3eff:fea7:6ad5
- Name in the model: not present
- Purpose: DNS server, not relevant for MeasrDroid's architecture.

**droid1** Virtual machine
- IPv4: 131.159.15.42
- IPv6: 2001:4ca0:2001:13:216:3eff:fe03:34f8
- Name in the model: *UploadDroid*
- Purpose: Receive data via http/https.

**c3po** Physical, powerful machine
- IPv4: 131.159.15.52
- IPv6: 2001:4ca0:2001:13:2e0:81ff:fee0:f02e
- Name in the model: *CollectDroid*
- Purpose: Trusted collection and storage.

The relevant machines are *UploadDroid* at 131.159.15.42 and *CollectDroid* at 131.159.15.52. We found that the machines do not have a firewall set up. All rely on the central firewall of our lab.

This central firewall may be the largest real-world, publicly available iptables firewall in the world and handles many different machines and networks. MeasrDroid is only a tiny fragment of it. We obtained a snapshot from June 2016 and make it publicly available [46]. The firewall is managed by several users and it consists of over 5500 rules.

[6] Except a management interface e.g. via SSH

MeasrDroid relies on the protocols http (port 80), https (port 443), and ssh (port 22). For conciseness, we focus our audit on port 80. Fundamentally, port numbers are not of a big concern for the overall architecture. Notably, our theoretical analysis, in particular Figure 3, has abstracted over concrete port number all the time.

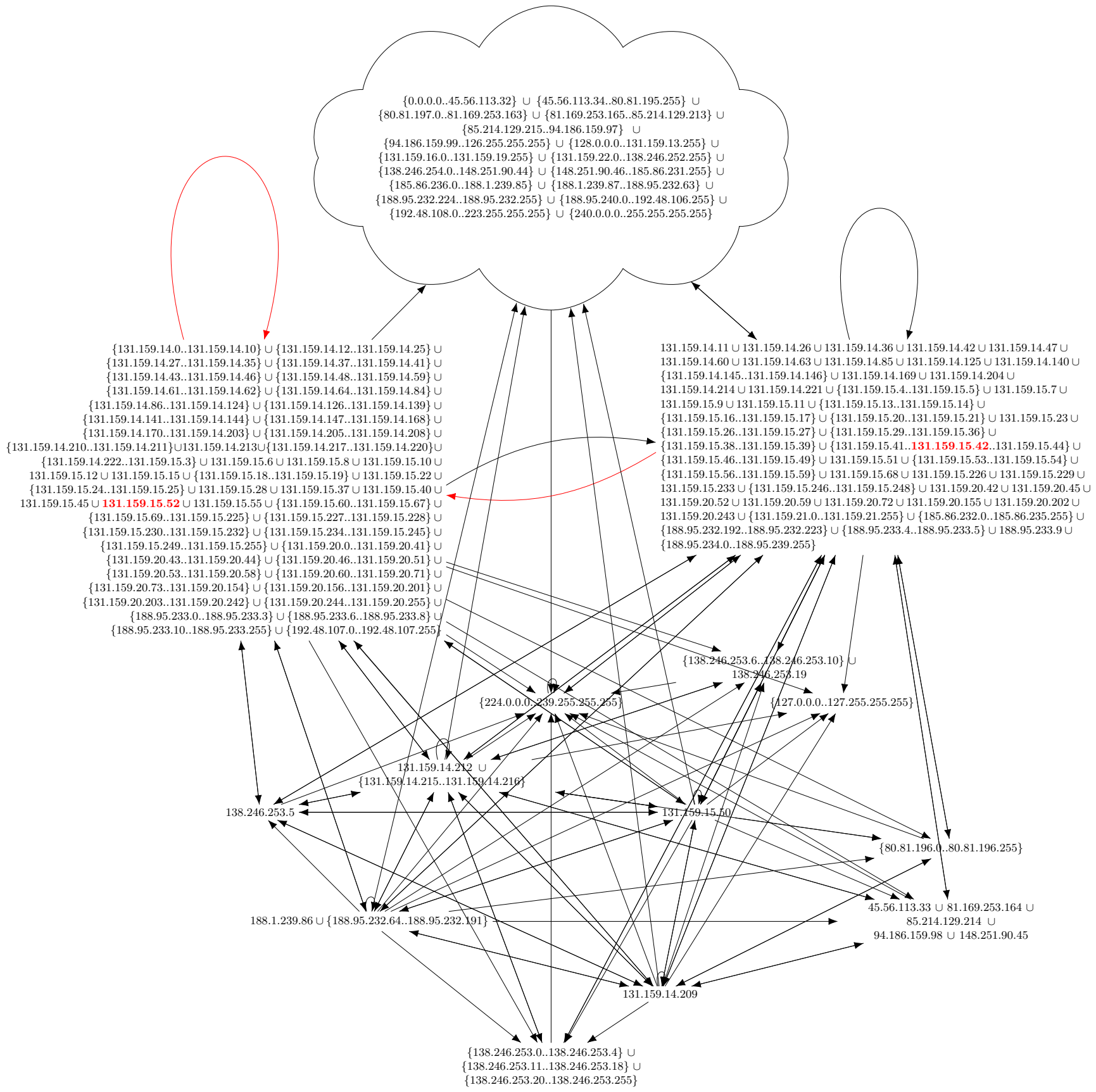

**Fig. 4.** MeasrDroid: Main firewall – IPv4 http connectivity matrix

The model of the MeasrDroid architecture (cf. Figure 3) should be recognizable in the rules of our central firewall. In particular, *CollectDroid* should not be reachable from the Internet, *UploadDroid* should be reachable from the Internet, and *CollectDroid* should be able to pull data from *UploadDroid*. This information may be hidden somewhere in the more than 5500 IPv4 firewall rules and over 6000 IPv6 firewall rules. We used the *fffuu* tool [47] to extract the access control structure of the firewall. The result is visualized in Figure 4 for IPv4. The IPv6 structure is shown in Figure 6. These figures may first appear highly confusing, which is due to the sheer intrinsic complexity of the access control policy enforced by the firewall. We have highlighted three entities in both figures. Because the structure and the results are similar for IPv4 and IPv6 and due to its long addresses the IPv6 graph is even worse readable than the IPv4 graph, we continue our analysis only with IPv4 for this paper. First, at the top, the IP range enclosed in a cloud corresponds to the IP range which is not used by our department, i.e. the Internet. The large block on the left corresponds to most internal machines which are not globally accessible. The IP address we marked in bold red in there belongs to *CollectDroid*. Therefore, inspecting the arrows, we have formally verified our first auditing goal: *CollectDroid* is not directly accessible from the Internet. The other large IP block on the right belongs to machines which are globally accessible. The IP address we marked in bold red in there belongs to *UploadDroid*. Therefore, we have verified our second auditing goal: *UploadDroid* should be reachable from the Internet. In general, it is pleasant to see that the two machines are in different access groups. Finally, we see that the class of IP addresses including *CollectDroid* can access *UploadDroid*. Therefore, we have verified our third auditing goal.

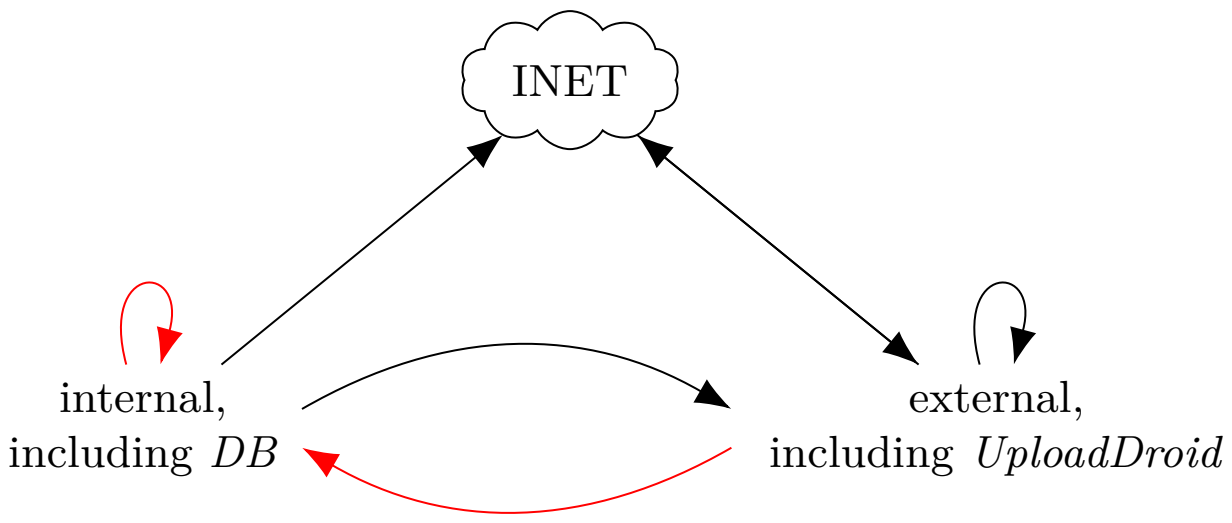

**Fig. 5.** MeasrDroid: Main firewall – simplified connectivity matrix

For the sake of example, simplicity, and presentiveness, we disregard that most machines at the bottom of Figure 4 could attack *CollectDroid*.[7] Therefore, the huge access control structure at the bottom of Figure 4 is not related to MeasrDroid and can be ignored. We have extracted only the relevant (and simplified) parts in Figure 5. In the previous paragraph, we only presented the successful parts of the audit. Our audit also revealed many problems related to MeasrDroid, visualized with red arrows. The problems can be clearly recognized in Figure 5:

[7] Our method is also applicable to the complete scenario; this would only decreases clarity without contributing any new insights. We acknowledge the sheer complexity of this real-world setup with all its side-constraints.

- *UploadDroid* can connect to *CollectDroid*. This is a clear violation of the architecture. We have empirically verified this highly severe problem by logging into *UploadDroid* and connecting to *CollectDroid*.
- In general, most internal machines may access *CollectDroid*, which violates the architecture.
- There are no restrictions for *UploadDroid* with regard to outgoing connections. In theory, it should only passively retrieve data and never initiate connections by itself (disregarding system updates).
- We uncovered a special IP address with special access rights towards *CollectDroid* (only shown in the full figure). We found an abandoned server, which has no current relevance for the MeasrDroid system. As a consequence, the access rights were revoked.

Therefore, our audit could verify some core assertions about the actual implementation. In addition, our audit could uncover and confirm serious bugs in the implementation. These bugs were unknown prior to our audit and we could only uncover them with the help of the process proposed in this paper.

*Automatically fixing bugs* To fix the problems our audit uncovered, we decided to install additional firewall rules at *CollectDroid*. *topoS* could automatically generate the rules for us. We detail the *topoS*-supported firewall configuration in the next section.

*A Firewall for C3PO* *topoS* can generate a global firewall for the complete MeasrDroid architecture. We filtered the output for rules which affect *CollectDroid*. Note that *topoS* generates a fully functional *stateful* firewall (cf. [39,48]) for us. For our architecture, *topoS* generated the two simple rules shown in Figure 7.

The first rule allows *CollectDroid* to connect to the *UploadDroid*. The second rule allows the *UploadDroid* to answer to such existing connections.

In cooperation with the authors of the MeasrDroid system, we manually extended the rules to further allow some core network services such as ICMP, DHCP, DNS, NTP, and SSH for remote management. In addition, we allow further outgoing TCP connections from *CollectDroid* (for example for system updates) but log those packets. The modified ruleset is illustrated in Figure 8.

We analyzed the modified firewall with *fffuu* to ensure that it still conforms to the overall policy. *fffuu* immediately verified that *CollectDroid* is no longer reachable from any machine (excluding localhost connection) over http. We also verified that we will not lock ourselves out from ssh access from our internal network. After this verification, the firewall was deployed to the real *CollectDroid* machine. Similarly, we implemented and deployed an IPv6 firewall.

## 6 Discussion & Future Work

In the previous section we presented a structured approach to apply our formalism to two real-world software architectures and their implementation.

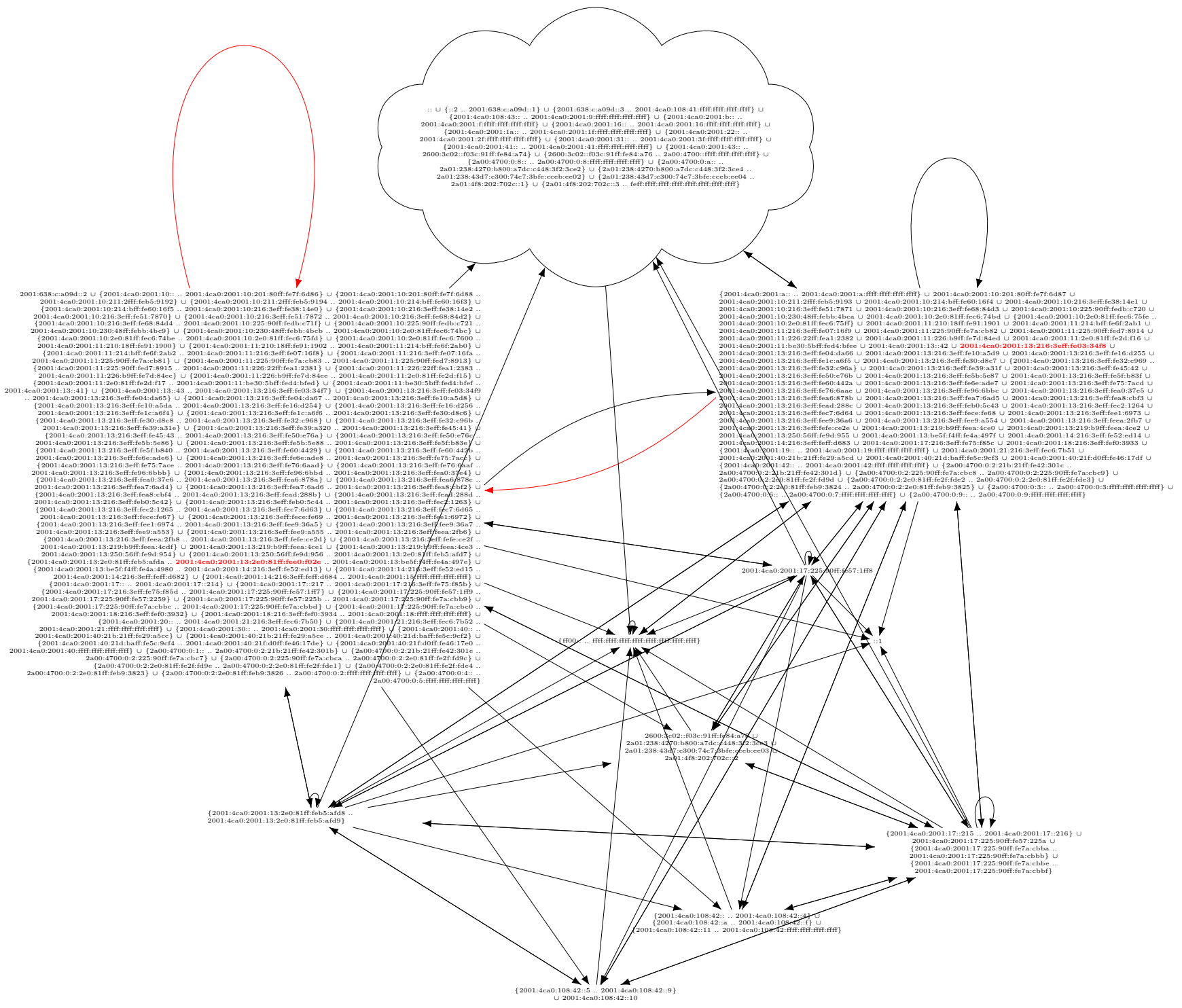

**Fig. 6.** MeasrDroid: Main firewall – IPv6 http connectivity matrix

```
*filter
:INPUT DROP [0:0]
:FORWARD DROP [0:0]
:OUTPUT DROP [0:0]
-A OUTPUT -s 131.159.15.52 -d 131.159.15.42 -j ACCEPT
-A INPUT -m state --state ESTABLISHED -s 131.159.15.42 -d 131.159.15.52 -j ACCEPT
COMMIT
```

**Fig. 7.** Automatically generated `iptables` rules

```
*filter
:INPUT DROP [0:0]
:FORWARD DROP [0:0]
:OUTPUT DROP [0:0]
# topoS generated:
# C3PO -> UploadDroid
-A OUTPUT -s 131.159.15.52 -d 131.159.15.42 -j ACCEPT
# UploadDroid -> C3PO (answer)
-A INPUT -m state --state ESTABLISHED -s 131.159.15.42 -d 131.159.15.52 -j ACCEPT
# custom additional rules
-A INPUT -i lo -j ACCEPT
-A OUTPUT -o lo -j ACCEPT
-A OUTPUT -p icmp -j ACCEPT
-A INPUT -p icmp -m state --state ESTABLISHED,RELATED -j ACCEPT
-A INPUT -s 131.159.20.190/24 -p tcp -m tcp --dport 22 -j ACCEPT
-A OUTPUT -m state --state ESTABLISHED -p tcp -m tcp --sport 22 -j ACCEPT
# DHCP
-A INPUT -p udp --dport 67:68 --sport 67:68 -j ACCEPT
-A OUTPUT -p udp --dport 67:68 --sport 67:68 -j ACCEPT
# ntp
-A OUTPUT -p udp --dport 123 -j ACCEPT
-A INPUT -p udp --sport 123 -j ACCEPT
# DNS
-A OUTPUT -p udp --dport 53 -j ACCEPT
-A INPUT -p udp --sport 53 -m state --state ESTABLISHED  -j ACCEPT
# further output, policy could be improved.
# Notice mails to admin, system updates, ...
-A OUTPUT -p tcp -j LOG
-A OUTPUT -p tcp -j ACCEPT
-A INPUT -p tcp -m state --state ESTABLISHED -j LOG
-A INPUT -p tcp -m state --state ESTABLISHED -j ACCEPT
COMMIT
```

**Fig. 8.** Manually tuned `iptables` rules

We could show that it is straightforward to derive a model from a real system and to make its most important aspects of privacy explicit: Participating systems, each consisting of individual components within a common boundary, their interconnections, and the classes of data they handle.

In the first case study, additionally the effect of privacy-enhancing practices (separation, aggregation, hiding) could be made visible. With the help of our model we assessed the effectiveness of these methods in the given scenarios. However, we also found that modeling methods of mere reduction of privacy-criticality could be improved by defining a partial order on the set of taint labels. Regarding privacy protection goals, it became easy to argue for (un)successful fulfillment of data minimization and unlinkability. The degree of achieving these goals becomes clear by visualizing the flows of critical data streams. In both studies, the mere creation of our system model improved the transparency, consequently allowing direct assessment of the previously mentioned goals. While this lays the foundation for intervenability, this goal is only achieved by sharing the created model with the subjects whose data is processed by the system. It will, however, depend on the service providers whether they do so.

For risk analysis, our model explicitly suggests components and systems for being assets, trusted parties, or attack targets. Furthermore, given their taint labels, it provides a foundation for rating their criticality. Here, especially components with privacy-enhancing functionality can be automatically identified by their non-empty untainting sets. Finally, in the second case study we could additionally show the automatic generation of remedies for the identified privacy violations. With help of *topoS* we could build firewall rules which enforce the model's architecture.

## 7 Conclusion

Several guidelines and policies for verifying and auditing the privacy properties of software architectures and IT systems exist. Yet, this task itself is usually carried out manually, making it complex and time-consuming.

In this paper, we presented a model, formalization, and technique based on static taint analysis and showed its applicability to the assessment of integral privacy protection goals. From given architecture designs or existing implementations, a model is derived which makes transmission and processing of privacy-critical data explicit. We showed how this information can be used to support the data protection goals of transparency, data minimality, and unlinkability.

We integrated our model into the formal policy framework *topoS* and proved soundness with Isabelle/HOL. Two real-world case studies demonstrate the applicability of our approach, exemplifying that insights formally derived from the model are consistent with manual inspections of the architecture. In the second studied system, auditing could be even carried out in a completely automated manner, uncovering previously unknown bugs, and providing measures for mitigation.

## Availability

Our formalization, case studies, and proofs can be found at `https://www.isa-afp.org/entries/Network_Security_Policy_Verification.shtml`

## Acknowledgments

This work has been supported by the German Federal Ministry of Education and Research, project DecADe, grant 16KIS0538.